\documentclass{article}
\usepackage{tgpagella}
\usepackage[T1]{fontenc}
\usepackage[utf8]{inputenc}
\usepackage{array}
\usepackage{booktabs}
\usepackage{bbding}
\usepackage{amsmath}
\usepackage{graphicx}

\makeatletter

\providecommand{\tabularnewline}{\\}

\@ifundefined{date}{}{\date{}}
\pdfoutput=1

\usepackage{subcaption}
\usepackage{listings}
\usepackage{color}
\usepackage{sourcecodepro}
\definecolor{highlight}{RGB}{192,80,77}
\usepackage[capitalise]{cleveref}
\usepackage[acronym]{glossaries}

\usepackage{smartdiagram}
\usesmartdiagramlibrary{additions}
\newcommand{\code}[1]{\lstinline[postbreak={}]{#1}}

\makeatother

\usepackage{listings}

\begin{document}

\title{User Manual for the \\
 Apple CoreCapture Framework}

\author{David Kreitschmann}

\maketitle

\newacronym{mDNS}{mDNS}{Multicast DNS} \newacronym{NVRAM}{NVRAM}{Non-Volatile
Random-Access Memory} \newacronym{TLV}{TLV}{Type-Length-Value}
\newacronym{WoWL}{WoWL}{Wake on Wireless LAN} 

CoreCapture is Apple's primary logging and tracing framework for IEEE
802.11 on iOS and macOS. It allows users and developers to create
comprehensive debug output for analysis by Apple. In this manual,
we provide an overview into the concepts, show in detail how CoreCapture
logs can be created on iOS and macOS, and introduce the first CoreCapture
dissector for Wireshark. 

\section{Motivation}

Debugging IEEE 802.11 issues often involves combining several sources
of information on the devices and often involves creating wireless
traces using a separate device, which may not be at hand in all scenarios.
Issues may only appear in very specific scenario, e.g. a specific
wireless network, thus it is critical for vendors to receive enough
information about an issue to fully understand it and provide a solution. 

To generate and package broad debug information Apple created the
CoreCapture framework. On demand or in case of driver errors CoreCapture
creates a folder with log files and dumps of internal hardware state,
such as a RAM dump of the Wi-Fi chip. The most interesting data for
us was found in PCAP trace files. These files combine driver communication
traces, log messages and network traces into one file.

During the log creation process accompanying information states that
``You will be able to review the log files on your computer prior
to sending them to Apple.'' \cite{Apple:ProfilesLogs}. Reviewing
these files is, unfortunately, not possible because there are no public
tools to read these proprietary files. Having tools at hand to read
these files gives users the possibility to make an informed decision
and protect their privacy. The generated data is tailored for the
analysis of driver and firmware issues by Apple but could be useful
for security researchers, e.g., for reverse engineering. Reverse engineering
the Apple Wireless Direct Link Protocol \cite{AWDLMobicom}, a proprietary
wireless ad hoc protocol, was our entry point to CoreCapture.

\section{Introduction to CoreCapture}

\begin{figure}
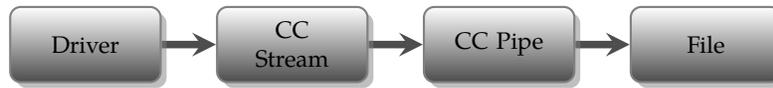

\begin{centering}
 \smartdiagramset{
  uniform color list=gray!60!black for 4 items,
  back arrow disabled=true,
  additions={
   additional item offset=0.85cm,
   additional item border color=red,
   additional connections disabled=false,
   additional arrow color=red,
   additional arrow tip=stealth,
   additional arrow line width=1pt,
   additional arrow style=]-latex’,
   }
}
\smartdiagram[flow diagram:horizontal]{Driver, CC Stream, CC Pipe, File} 
\par\end{centering}
\caption{CoreCapture flow of log entries\label{fig:CoreCapture-flow}}
\end{figure}

Traditionally creating debug logs, e.g., with syslog, starts with
increasing the amount of log data (e.g., by starting an application
in debug mode) and filling the disk with logs. This is not ideal for
mobile devices because issues may only appear very seldom and storage
is limited. In CoreCapture, information written to a \emph{stream}
that is stored temporarily in a kernel buffer (\emph{pipe}) until
an error condition is reached or a dump is triggered manually. Only
then a CoreCapture output is generated and saved to disk, as shown
in \cref{fig:CoreCapture-flow}. Therefore, creating a log in CoreCapture,
involves multiple steps:
\begin{enumerate}
\item Preparation:
\begin{enumerate}
\item increase the debug output generation (CoreCapture stream)
\item configure CoreCapture to log continuously (CoreCapture pipe)
\end{enumerate}
\item Reproduce the issue or condition
\item Save the logs
\begin{enumerate}
\item dump the log to persistent storage
\item transfer logs to a computer (iOS/watchOS only)
\end{enumerate}
\end{enumerate}

\section{Creating CoreCapture traces on iOS}

Apple provides instructions on their developer website on the creation
of logs for bug reports \cite{Apple:ProfilesLogs}.

\begin{figure}
\begin{centering}
\includegraphics[scale=0.28]{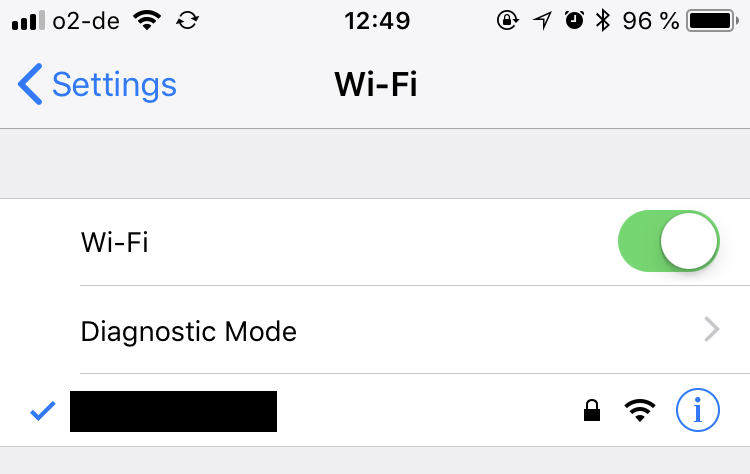}
\par\end{centering}
\caption{iOS Wi-Fi Diagnostic Mode in Wi-Fi Settings\label{fig:iOS-Wi-Fi-Diagnostic}}
\end{figure}

\begin{figure}
\begin{centering}
\includegraphics[scale=0.28]{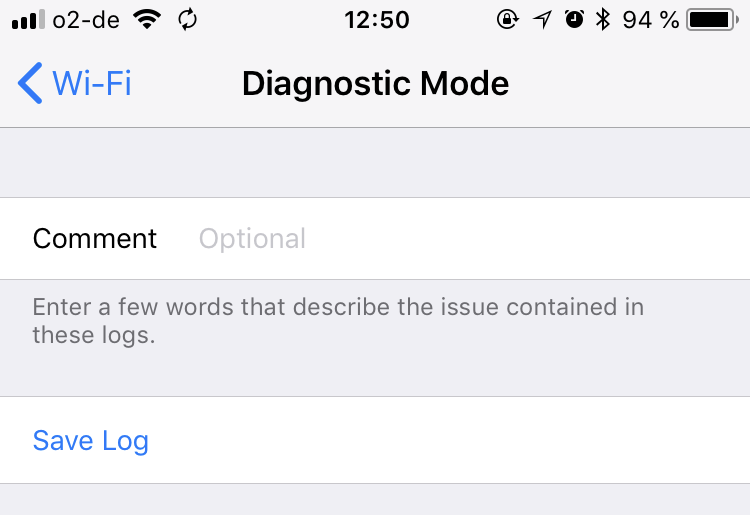}
\par\end{centering}
\caption{iOS Wi-Fi Diagnostic Mode\label{fig:iOS-Wi-Fi-Diagnostic-1}}
\end{figure}

In the case of Wi-Fi logs Apple provides a cryptographically signed
configuration profile (\code{MegaWifiProfile.mobileconfig}) to enable
the Wi-Fi diagnostics mode on iOS (and even watchOS) devices. The
profile is set to expire after 31 days, therefore this process works
only as long as Apple provides updated configuration profiles. To
install the configuration profile, the file must be opened on an iOS
device (e.g., by opening the link on the device or sending the file
via mail or AirDrop). After the profile is installed, a new menu item
with a \emph{Save Log} button is available in the Settings app (\cref{fig:iOS-Wi-Fi-Diagnostic}
and \ref{fig:iOS-Wi-Fi-Diagnostic-1}).

\begin{figure}
\begin{centering}
\includegraphics[scale=0.21]{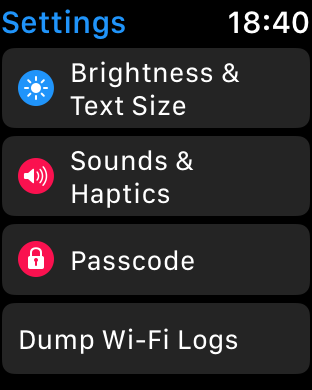}
\par\end{centering}
\caption{watchOS ``Dump Wi-Fi Logs'' button\label{fig:watchOS-Dump-Wi-Fi}}
\end{figure}

After clicking the \emph{Save Log} button, capture files are saved
to flash storage, but are not visible in the iOS UI. They must be
exported to a connected computer. The officially supported way to
export the files from iOS is through iTunes \cite{Apple:ProfilesLogs}.
When syncing with iTunes on the Mac, CoreCapture files are copied
to the folder \code{/Users/<USER>/Library/Logs/CrashReporter/MobileDevice/<Device
Name>/CoreCapture}.

If no Mac or Windows computer is available, we can also copy these
files using the \code{idevicecrashreport} tool from the \emph{libimobiledevice}
\cite{libimobiledevice} suite.

\subsection{Creating traces on watchOS}

The process on watchOS is almost the same as on iOS: When opening
the configuration profile on an iPhone a user can choose to install
it on a connected Apple Watch. The trigger menu item then appears
directly in the Settings App on the Watch (\cref{fig:iOS-Wi-Fi-Diagnostic}). 

While the Watch is charged, traces are transferred to the connected
iPhone. Afterwards they can be copied to a computer, just as with
iOS. The files can be found in \code{/Users/<USER>/Library/Logs/CrashReporter/Mobile
Device/<DeviceName>/ProxiedDevice-<UUID>/CoreCapture}.

\subsection{More details on the MegaWifi configuration profile.}

Configuration profiles are based on Apple's Property List files an
optional signature wrapper. Therefore they can be read with a standard
text editor. The profile contains three different payloads: 
\begin{description}
\item [{com.apple.corecapture.configure}] Sets CoreCapture parameters,
as shown exemplarily in \cref{fig:iOS-MegaWifi-Excerpt}. This, e.g.,
shows that the \emph{FirmwareLogs/Chip\_UART} logstream is set to
loglevel 5 and LogFlags is set to 1. The FirmwareLogs pipe's policy
is set to 1 (=continuous). This is replicated very similarly for other
Wi-Fi drivers and streams. Unfortunately there is no documentation
which loglevel and flag parameters are available.
\item [{com.apple.system.logging}] Increased log output from several Wi-Fi
related processes. This output is sent to Apple's Unified Logging
framework and is available in the Console application on a Mac. Unified
Logging is not part of this manual.
\item [{com.apple.defaults.managed}] Enables the diagnostic menu item,
increases log output for several processes and adds configuration
redundant to the payload of \emph{com.apple.corecapture.configure}. 
\end{description}
\begin{figure}
\inputencoding{latin9}\begin{lstlisting}[language=XML]
<key>CCConfigurePipe</key>
<dict>
	<key>com.apple.driver.AppleBCMWLANCoreV3.0</key>
	<dict>
		<key>FirmwareLogs</key>
		<dict>
			<key>Policy</key>
			<integer>1</integer>
		</dict>
	</dict>
</dict>
<key>CCConfigureStream</key>
<dict>
	<key>com.apple.driver.AppleBCMWLANCoreV3.0</key>
	<dict>
		<key>FirmwareLogs</key>
		<dict>
			<key>Chip_UART</key>
			<dict>
				<key>CoreCapture</key>
				<dict>
					<key>LogFlags</key>
					<integer>1</integer>
					<key>LogLevel</key>
					<integer>5</integer>
				</dict>
			</dict>
		</dict>
	</dict>
</dict>
\end{lstlisting}
\inputencoding{utf8}
\caption{iOS MegaWifiProfile excerpt\label{fig:iOS-MegaWifi-Excerpt}}
\end{figure}

As stated above the profile is signed by Apple, which is strictly
only required for installing the\emph{ com.apple.defaults.managed}
payload, especially enabling the otherwise hidden menu item. Therefore
a custom unsigned configuration profile to configure CoreCapture can
be created and installed, but it is not possible to enable the hidden
menu item. If two profiles are installed, Apple's profile seems to
take precedence over an unsigned profile, making it impossible to
override e.g., loglevels.

It may also be possible to trigger CoreCapture by inducing an error
in the driver, because in some conditions CoreCapture is triggered
automatically, but this is outside the scope of this manual.

\section{Creating traces on macOS}

Apple's official instructions \cite{Apple:ProfilesLogs} to create
Wi-Fi logs on the Mac involve using the Wireless Diagnostics Application\footnote{alt-click the Wi-Fi symbol in the menu bar to run Wireless Diagnostics},
which amongst other information also triggers a manual capture. Unfortunately,
the timing of these captures can't be influenced so it may not be
ideal for all cases. Therefore, a better way is to set up CoreCapture
manually.

It is possible to install the iOS configuration profile on a Mac,
but this provides only logging for the shared part of the Wi-Fi driver
and no manual trigger in the UI. But this is not required anyway,
because Apple provides us with the \code{cctool} command line tool,
deep inside the CoreCapture Framework.\footnote{/System/Library/PrivateFrameworks/CoreCaptureControl.framework/Resources/cctool}
It can be used to change settings, just as the configuration profile,
and trigger a manual capture. Unfortunately is not possible to list
all available \emph{pipes} and \emph{streams} with \code{cctool},
but the information is available through another framework: IOKit.

\subsection{IOKit's IORegistry}

\begin{figure}
\centering \includegraphics[width=0.7\linewidth]{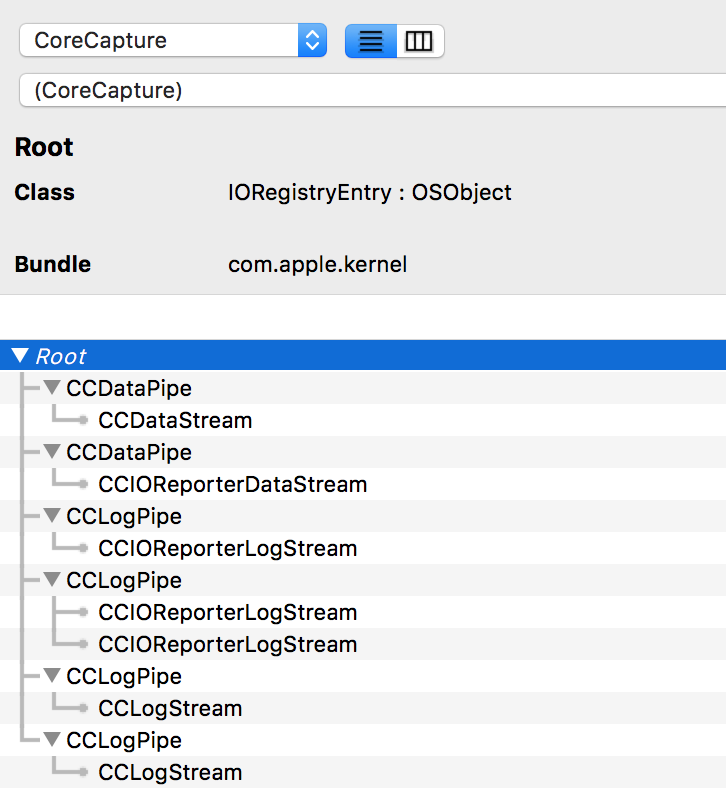}
\caption{CoreCapture Pipes and LogStreams}
\label{fig:CoreCapturePipesLogs} 
\end{figure}

\begin{figure}
\centering \includegraphics[width=0.7\linewidth]{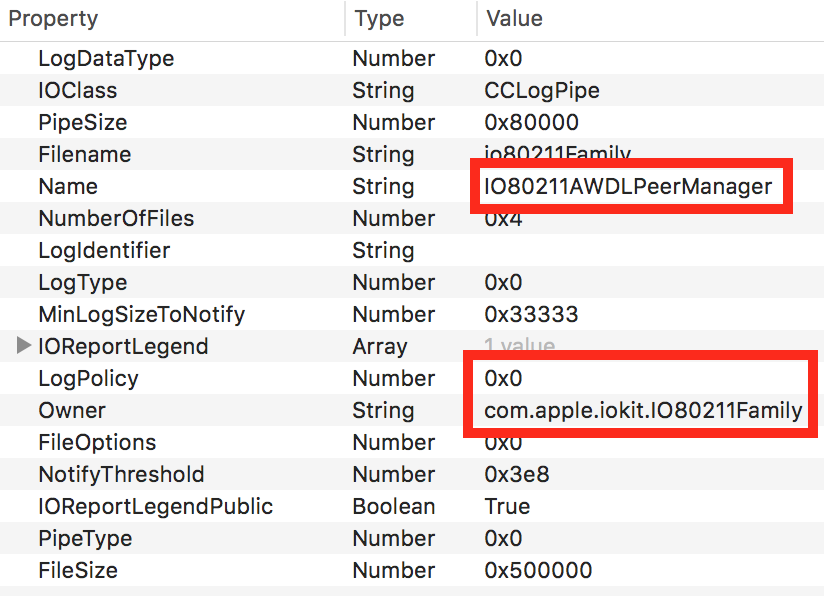}
\caption{CoreCapture Pipe properties}
\label{fig:CoreCapture-Pipe} 
\end{figure}

IOKit includes the I/O Catalog which is responsible for matching a
device to a driver and the tree-based I/O Registry where all actual
devices and properties are listed. It is also the place to find the
available CoreCapture pipes and streams. The IORegistry can be queried
with \emph{IORegistryExplorer}\footnote{Part of Apples \emph{Additional Tools for Xcode}}
or the \code{ioreg} command. When switching to the CoreCapture
mode in \emph{IORegistryExplorer} (\cref{fig:CoreCapturePipesLogs}),
several \emph{pipe }objects with \emph{stream} objects underneath
are shown. Pipes can be understood as internal buffers and have, amongst
others, a \code{LogPolicy}, \code{Name}, and \code{Owner}
(\cref{fig:CoreCapture-Pipe}). The stream objects underneath belong
to the pipe. Therefore, they are addressed with the \code{Name}
and \code{Owner} of the pipe as well as their own \code{Name}.

Different types of streams can be found here. A \emph{LogStream} is
a configurable information source, such as the logging facility of
a driver. It can be configured as needed with loglevels, steering
to the amount of information, and flags, setting which subsystems
to be logged (\cref{fig:CoreCapture-streams}). A \emph{DataStream}
is used for information generated once during a capture. This includes
snapshots of the internal state of the driver, such as the RAM of
the Wi-Fi chip.

\subsection{Using cctool}

cctool provides some help with the \code{-h} parameter and using
the parameters together with the information found in the I/O Registry,
CoreCapture properties can be changed:

\texttt{}\inputencoding{latin9}
\begin{lstlisting}[language=bash]
sudo ./cctool -o com.apple.driver.AirPort.Brcm4360.0 -p DriverLogs -x 1
\end{lstlisting}
\inputencoding{utf8}
In this case, the parameter \code{-x 1} sets the policy to \emph{continuous},
which can also be verified using the I/O Registry.

\begin{figure}
\centering \includegraphics[width=0.7\linewidth]{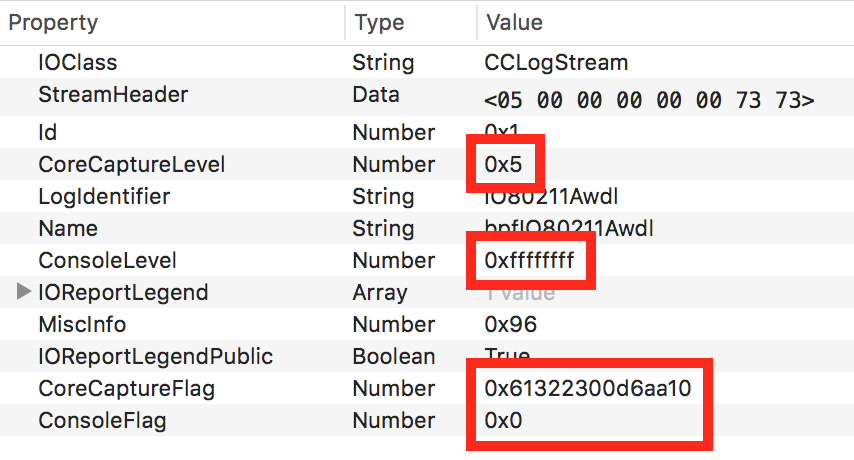}
\caption{CoreCapture LogStream properties}
\label{fig:CoreCapture-streams} 
\end{figure}

The process is roughly the same for the \emph{LogStream} objects
(\cref{fig:CoreCapture-streams}): \code{cctool} is used to change
log levels and log flags and they can be set differently for console
and CoreCapture logs. Unfortunately, there is no list of the available
loglevels and flags for macOS, but most default values from the iOS
configuration profile can also be adapted for macOS. Values can be
set to the maximum value with \code{-1}, resulting in, e.g., \code{0xffffffff}.

A full example for the Broadcom driver and \code{IO80211PeerManager}
(the main class handling AWDL in \code{IO80211Family}) is presented
in \cref{fig:cctool-example}. Capture files are saved to \code{/Library/Logs/CrashReporter/CoreCapture}.

\begin{figure}
 \inputencoding{latin9}
\begin{lstlisting}[language=bash]
export CCTOOL=/System/Library/PrivateFrameworks/CoreCaptureControl.framework/Resources/cctool
sudo $CCTOOL -o com.apple.driver.AirPort.Brcm4360.0 -p DriverLogs -x 1
sudo $CCTOOL -o com.apple.driver.AirPort.Brcm4360.0 -p DriverLogs -s DriverLogs -l 5 -f 8388608 
sudo $CCTOOL -o com.apple.iokit.IO80211Family -p IO80211AWDLPeerManager -x 1
sudo $CCTOOL -o com.apple.iokit.IO80211Family -p IO80211AWDLPeerManager -s bpfIO80211Awdl -l 5 -f 27358198660246032 -g 1 -m 0
sudo $CCTOOL -o "*" -p "*" -c manual_dump 
\end{lstlisting}
\inputencoding{utf8} \caption{cctool commands using log settings from Apples iOS profile\label{fig:cctool-example}}
\end{figure}

\subsection{Increasing driver output on macOS further}

\label{sec:debug-boot-args}

More debug information is available in macOS by enabling additional
flags in the boot arguments. The boot arguments need to be present
in the \gls{NVRAM} when the system boots. Thus, this is not possible
with unrooted iOS devices. On macOS, access to the \gls{NVRAM}
is blocked by \emph{System Integrity Protection} (SIP). This can be
circumvented by booting the computer in recovery mode by pressing
\code{cmd+R} during a reboot. The terminal can be accessed in the
menu bar and the commands listed in \cref{fig:nvram} can be applied.
These enable very verbose information in CoreCapture, although they
could probably be set more selectively.

We note that disabling part of SIP which implements key security features
of macOS renders the system vulnerable to attacks. Thus, we recommend
to only do this on a dedicated and non-production machine.

\begin{figure}
\texttt{}\inputencoding{latin9}
\begin{lstlisting}[language=bash]
csrutil enable --without nvram
nvram boot-args=debug=0x10000 awdl_log_flags=0xffffffffffffffff awdl_log_flags_verbose=0xffffffffffffffff awdl_log_flags_config=1 wlan.debug.enable=0xff
\end{lstlisting}
\inputencoding{utf8}
\caption{Disabling part of the\protect\emph{System Integrity Protection} and
setting the \code{boot-args} NVRAM variable\label{fig:nvram}}
\end{figure}

\section{A CoreCapture trace folder in detail}

\begin{figure}
\begin{centering}
\includegraphics[scale=0.5]{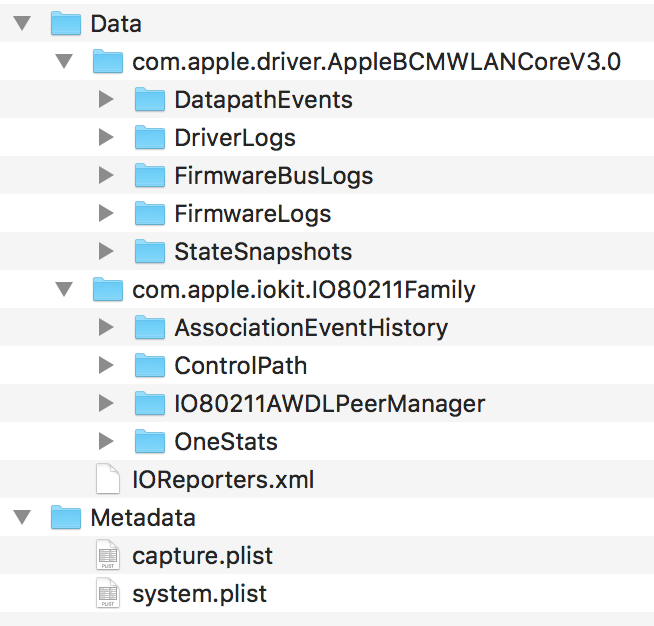}
\par\end{centering}
\caption{iOS CoreCapture folder}
\label{fig:ios-corecapture-folder}

\end{figure}

An exemplary capture is shown in \cref{fig:ios-corecapture-folder}.
The \code{Metadata} folder holds information about the system and
the time of the capture. The data folder holds the \code{com.apple.iokit.IO80211Family}
folder with output from Apple's higher level IEEE 802.11 stack and
a adapter-specific folder such as \code{com.apple.driver.AppleBCMWLANCoreV3.0}
with more device-specific output. This mirrors the structure of the
pipes and streams in the I/O Registry. \cref{tab:cc-files} shows
the content of these folders and their file format.

\begin{table}
\begin{tabular}{>{\raggedright}p{0.25\columnwidth}>{\raggedright}p{0.35\columnwidth}>{\raggedright}p{0.08\columnwidth}ll}
\toprule 
Name & Content & Format & iOS & macOS\tabularnewline
\midrule
\midrule 
\multicolumn{5}{l}{com.apple.driver.AppleBCMWLANCore/com.apple.driver.Brcm4360}\tabularnewline
\midrule
\midrule 
DatapathEvents & Traces for all network interfaces

IOCTLs between IO80211Family and the Broadcom driver

AWDL specific output & PCAP & \Checkmark{} & \tabularnewline
\midrule 
DriverLogs & Broadcom driver logs & TXT & \Checkmark{} & \Checkmark{}\tabularnewline
\midrule 
FirmwareBusLogs & unidentified traces & PCAP & \Checkmark{} & \tabularnewline
\midrule 
FirmwareLogs & Broadcom firmware logs & TXT & \Checkmark{} & \tabularnewline
\midrule 
StateSnapshots & internal state and memory dumps & TXT, BIN & \Checkmark{} & \Checkmark{}\tabularnewline
\midrule
\midrule 
\multicolumn{5}{l}{com.apple.iokit.IO80211Family}\tabularnewline
\midrule 
AssociationEvent History & unidentified content, often empty & XML & \Checkmark{} & \Checkmark{}\tabularnewline
\midrule 
ControlPath & IOCTLs and Events between user space and IO80211Family & PCAP & \Checkmark{} & \Checkmark{}\tabularnewline
\midrule 
IO80211AWDL PeerManager & AWDL Traces, including log messages & PCAP & \Checkmark{} & \Checkmark{}\tabularnewline
\midrule 
OneStats & I/O Registry output & XML & \Checkmark{} & \Checkmark{}\tabularnewline
\midrule 
IOReporters.xml & I/O Registry output & XML & \Checkmark{} & \Checkmark{}\tabularnewline
\bottomrule
\end{tabular}

\caption{Overview of the files inside a CoreCapture folder \label{tab:cc-files}}
\end{table}

\section{Wireshark Dissector}

\label{sec:cc_wireshark}Wireshark is a modular packet analyzer and
uses the PCAP format natively. Therefore, we decided to extend Wireshark
with a CoreCapture dissector. We provide our CoreCapture dissector
as part of our Wireshark fork at \cite{awdl-wireshark}.

\subsection{Building Wireshark}

Wireshark can be compiled as described in the included Readme file.
On macOS we recommend using the homebrew package manager \cite{homebrew}:

\inputencoding{latin9}\begin{lstlisting}[language=bash]
brew install --only-dependencies --with-qt wireshark
git clone <git url> wireshark-awdl # or download an archive
mkdir wireshark-build
cd wireshark-build
export PATH="/usr/local/opt/qt/bin:$PATH"
cmake ../wireshark-awdl
make
open run/Wireshark.app
\end{lstlisting}
\inputencoding{utf8}

\subsection{Creating DLT\_USER Mapping}

\begin{figure}
\begin{centering}
\includegraphics[scale=0.4]{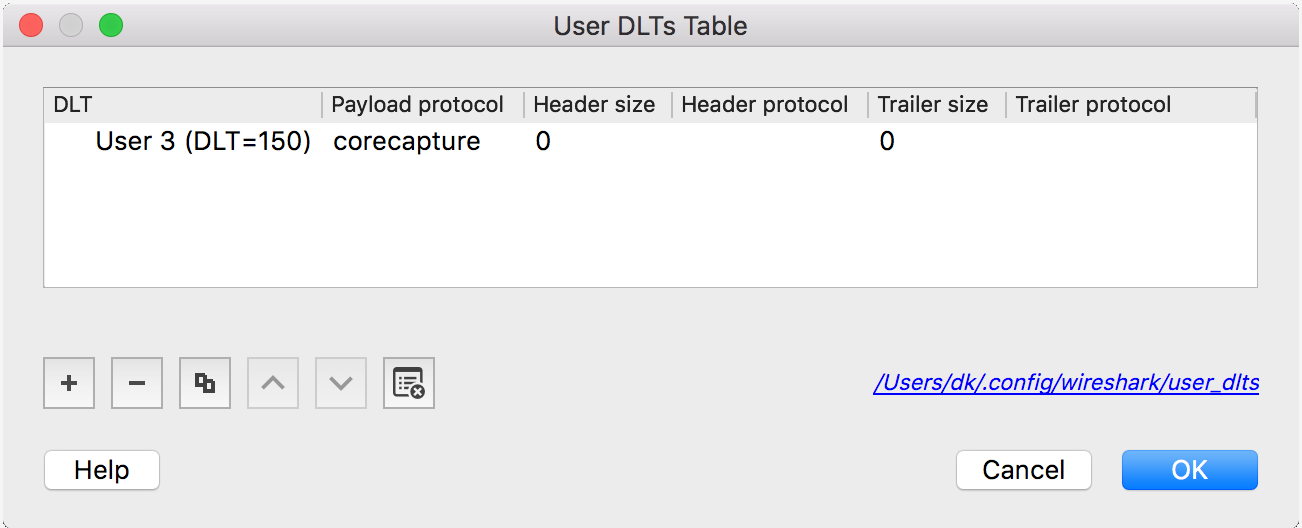}
\par\end{centering}
\caption{Wireshark User DLTs Table\label{fig:Wireshark-User-DLTs}}

\end{figure}

There is one last step to do before Wireshark can decode CoreCapture
files successfully, namely, creating a mapping to the CoreCapture
dissector. PCAP files are a container format for different types traces,
such as 802.11, Bluetooth, Ethernet IP. A link-layer type value included
in the PCAP file at creation time indicates at which layer to start
the frame dissection. Apple chose to use their own proprietary format
and non-unique link-layer type reserved for private use instead of
existing link layer types \cite{Tcpdump:Linktypes} or requesting
a unique identifier. Fortunately Wireshark allows to associate a custom
dissector with these types:

Open the Wireshark Preferences select \emph{Protocols $\rightarrow$ DLT\_USER}
and Edit the Encapsulation table. Add User 3 (DLT=150) to the table
with \code{corecapture} as payload protocol, as shown in \cref{fig:Wireshark-User-DLTs}.

\pagebreak{}

\bibliographystyle{IEEEtran}
\bibliography{references}

\end{document}